\documentclass[twocolumn,reprint,aps,prl, floatfix,amsmath,amssymb,superscriptaddress]{revtex4-2}

\usepackage[utf8]{inputenc}  
\usepackage{graphicx}
\usepackage{bm}
\usepackage{braket}
\usepackage{xcolor} 
\usepackage{subfigure}     
\usepackage{diagbox}
\usepackage{amsthm}
\usepackage{hyperref}

\definecolor{dkgreen}{rgb}{0,0.6,0}

\definecolor{purple}{rgb}{0.5,0,0.5}

\begin{document}
	
	\title{Thermodynamics of Schwarzschild-AdS black hole in non-commutative geometry}
	
	\author{Rui-Bo Wang}\author{Shi-Jie Ma}\author{Lei-You}\author{Jian-Bo Deng}
	\email[Jian-Bo Deng: ]{dengjb@lzu.edu.cn}\author{Xian-Ru Hu}
	\affiliation{Lanzhou Center for Theoretical Physics, Key Laboratory of Theoretical Physics of Gansu Province, Lanzhou University, Lanzhou, Gansu 730000, China}
	
	\date{\today}

    \begin{abstract}
	In this paper, we study the thermodynamics of Schwarzschild-anti-de Sitter black holes within the framework of non-commutative geometry. By solving the Einstein's equations, we derive the corrected Schwarzschild-AdS black hole with Lorentzian distribution and analyze the thermodynamics. Our results confirm that if the energy-momentum tensor outside the event horizon is related to the mass of the black hole, the conventional first law of thermodynamics will be violated. The study of criticality reveals that the black hole undergoes a small black hole-large black hole phase transition similar to that of the Van der Waals system, with a critical point and a critical ratio slightly smaller than that of the Van der Waals fluid. As the non-commutative parameter increases, the phase transition process shortens, leading to a critical point, and ultimately to the disappearance of the phase transition. The violation of the conventional first law results in a discontinuity of the Gibbs free energy during the phase transition, indicating the occurrence of zeroth-order phase transition. Moreover, we investigate the Joule-Thomson expansion, obtaining the minimum inversion temperature and the minimum inversion mass.

\end{abstract}

\maketitle
\section{Introduction}\label{Sect1}
Non-commutative geometry is a theory of spacetime quantization~\cite{noncomu1,noncomu2,noncomu3,noncomu4,noncomu5,noncomu6,noncomu7}, described by commutator $\left[\hat{x}^{\mu},\hat{x}^{\nu}\right]=\mathrm{i}\hat{\Theta}^{\mu\nu}$, where $\hat{x}^{\mu}$ and $\hat{x}^{\nu}$ are spacetime coordinate operators, $\hat{\Theta}^{\mu\nu}$ is an anti-symmetric constant tensor, $\Theta$ is a parameter with dimension of $\left[\mathrm{L}^{2}\right]$ that represents the minimum spacetime scale, and $\sqrt{\Theta}$ is a small quantity of the order of the Planck length $\ell_{p}$~\cite{commutator,planck-length}. Extensive research has been conducted to consider the effects of non-commutative fields in gravitational theories~\cite{gr1,gr2,gr3,gr4,gr5,gr6,gr7,gr8}. An interesting fact is that the commutator of coordinate operators being a nonzero constant $\Theta$ matrix is incompatible with Lorentz covariance~\cite{noncomu1}, which potentially suggests a close relationship between non-commutative fields and the violation of Lorentz symmetry. On the other hand, this also provides a new insight into theories about the violation of Lorentz covariance~\cite{lorentz1,lorentz2,lorentz2,lorentz3,lorentz4,lorentz5,lorentz6}. Non-commutative geometry also provides a reasonable modification to Schwarzschild spacetime. Compared to the point-like mass distribution in the ordinary Schwarzschild spacetime, the mass distribution of a spherically symmetric star is altered to a Gaussian distribution $\rho=\frac{M}{\left(4\pi\Theta\right)^{\frac{3}{2}}}\exp\left(-\frac{r^{2}}{4\Theta}\right)$ and a Lorentzian distribution $\rho=\frac{M\sqrt{\Theta}}{\pi^{\frac{3}{2}}\left(r^{2}+\pi\Theta\right)^{2}}$ spread throughout the entire space~\cite{distribution1}. Unlike the Schwarzschild spacetime with singularities, these two distributions both lead to a regular de Sitter core at short distances. There have been many studies investigating the properties of black holes (BHs) in the background of non-commutative geometry~\cite{distribution1,gr1,gr2,gr3,gr4,Gaussianthermo,optics,thermo1,thermo2,thermo3}. In~\cite{gr1}, the authors found that the evaporation endpoint of a Schwarzschild BH, corrected by a Gaussian distribution, is identified as a zero-temperature extremal BH, even in the case of electrically neutral and non-rotating objects. Moreover, there exists a finite maximum temperature that the BH can attain before eventually cooling down to absolute zero. The Schwarzschild BH with the Lorentzian distribution has been extensively studied as well, including its gravitational lensing effects, quasinormal modes, BH shadows, and thermodynamic properties~\cite{optics}.

BH thermodynamics is a relatively young branch of BH physics. Since Hawking and Page made the groundbreaking discovery of phase transitions in Schwarzschild-AdS BHs in 1983~\cite{Hawking1983}, this area has garnered significant attention and research, particularly concerning thermodynamics in anti-de Sitter (AdS) spacetime, which often yields more intriguing and profound conclusions. By correlating the BH's mass and the cosmological constant with the enthalpy and the pressure, researchers derived the first law of AdS-BH thermodynamics~\cite{firstlaw}, laying the groundwork for a detailed exploration of this subject in AdS spacetime. A classic example is the Reissner-Nordstr$\mathrm{\ddot{o}}$m (RN)-AdS BH, whose thermodynamic properties are thoroughly researched in~\cite{RNcriticality}, revealing phase transitions akin to those of the Van der Waals fluids, alongside critical points characterized by identical critical ratios and critical exponents. Additionally, the RN-AdS BH undergoes a Joule-Thomson process with inversion points, accompanied by a minimum critical mass~\cite{JT1}. Overall, extensive research indicates that BHs in AdS spacetime exhibit thermodynamic properties similar to those of the Van der Waals systems~\cite{RNcriticality,thermo_1,thermo_2,thermo_3,thermo_4,thermo_5,thermo_6,maxareats}. Notably, for a static spherically symmetric BH, if the energy-momentum tensor of the spacetime explicitly includes the BH's mass, the first law of thermodynamics for BHs requires modification~\cite{thermo_4,thermo_5,rev2}, complicating the study of enthalpy and Gibbs free energy.

In this work, we investigate the thermodynamics of the Schwarzschild-AdS BHs with Lorentzian distribution in non-commutative geometry. The article is organized as follows. In Sect.~\ref{Sect2}, we substitute the Lorentzian distribution into the Einstein equation with the cosmological constant to obtain the metric of Schwarzschild-AdS BH in non-commutative geometry, and discuss the existence of BH solutions. In Sect.~\ref{Sect3}, we conduct a detailed research on thermodynamics, including the revised first law, the equation of state, the behavior of phase transitions and criticality. We calculate the heat capacity and Gibbs free energy of the BH as well. In Sect.~\ref{Sect4}, we investigate the Joule-Thomson process. We study the inversion points and determine the conditions for their existence, specifically the minimum inversion temperature and minimum inversion mass. Finally, we summarize our findings and give a outlook in Sect.~\ref{Sect5}. In this paper, the Planck units $\hbar=k_{B}=G=c=1$ are used.

\section{Schwarzschild-AdS BH in non-commutative geometry}\label{Sect2}
The Einstein equation with the cosmological constant $\Lambda$ is
\begin{equation}\label{eq2_1}
	R_{\mu}^{\nu}-\frac{1}{2}\delta_{\mu}^{\nu}R+\delta_{\mu}^{\nu}\Lambda=8\pi T_{\mu}^{\nu},
\end{equation}
where $R_{\mu}^{\nu}$ is Ricci tensor, $\delta_{\mu}^{\nu}$ is Kronecker symbol, $R$ is Ricci scalar and $T_{\mu}^{\nu}$ is energy-momentum
tensor. A spherically symmetric spacetime could be expressed as
\begin{equation}\label{eq2_2}
	\mathrm{d}s^{2}=-f\left(r\right)\mathrm{d}t^{2}+f\left(r\right)^{-1}\mathrm{d}r^{2}+r^{2}\mathrm{d}\theta^{2}+r^{2}\sin^{2}\theta\mathrm{d}\phi^{2}.
\end{equation}
By substituting the metric into the Einstein equation, one could obtain
\begin{equation}\label{eq2_3}
	f\left(r\right)=1+\frac{1}{r}\int_{0}^{r} \left(8\pi r^{2} T_{0}^{0}-r^{2}\Lambda\right)\mathrm{d}r.
\end{equation}
The Lorentzian distribution $\rho$ of mass density in non-commutative geometry is~\cite{distribution1,optics,gr1,Gaussianthermo}
\begin{equation}\label{eq2_4}
	T_{0}^{0}=-\rho=-\frac{M\sqrt{\Theta}}{\pi^{\frac{3}{2}}\left(r^{2}+\pi\Theta\right)^{2}},
\end{equation}
where $M$ is the mass of the BH and $\Theta$ is the non-commutative parameter with dimension of $\left[\mathrm{L}^{2}\right]$. The metric of Schwarzschild-AdS BH in non-commutative geometry is
\begin{equation}\label{eq2_5}
	f\left(r\right)=1-\frac{2M}{r}+\frac{8M\sqrt{\Theta}}{\sqrt{\pi}r^{2}}-\frac{\Lambda r^{2}}{3}+\mathcal{O}\left(\Theta^\frac{3}{2} \right).
\end{equation}
It is clear that the correction of non-commutative geometry is of higher order.

In order to simplify the calculations, we introduce a parameter $a$ with dimension of $\left[\mathrm{L}\right]$
\begin{equation}\label{eq2_6}
	a=\frac{8\sqrt{\Theta}}{\sqrt{\pi}}.
\end{equation}
$a$ and $\Theta$ are equivalent, as they both can be used to characterize the strength of the noncommutative geometry.

The metric of a Schwarzschild-AdS BH in non-commutative geometry is given by
\begin{equation}\label{eq2_7}
	f\left(r\right)=1-\frac{2M}{r}+\frac{aM}{r^{2}}-\frac{\Lambda r^{2}}{3}.
\end{equation}
To ensure the existence of BH, it is necessary to satisfy $f\left(r_{h}\right)=0$ ($r_{h}$ is the radius of event horizon), $M>0$ and $r_{h}>0$. And it is obvious that $a>0$. On the other hand, AdS spacetime meets with $\Lambda<0$. These conditions impose these following constraints
\begin{equation}\label{eq2_8}
	M>a,~~~r_{h}>\frac{a}{2}.
\end{equation}

\section{Thermodynamics}\label{Sect3}

\subsection{The first law and Smarr relation}\label{Sect3_1}
The BH's mass, also regarded as its enthalpy, can be solved from $f\left(r_{h}\right)=0$,
\begin{equation}\label{eq3_1_1}
	M=\frac{-3r_{h}^{2}+\Lambda r_{h}^{4}}{3a-6r_{h}}.
\end{equation}
The temperature of the BH is defined by its surface gravity
\begin{equation}\label{eq3_1_2}
	T=\frac{f'\left(r_{h}\right)}{4\pi}=\frac{3a-3r_{h}-2a\Lambda r_{h}^{2}+3\Lambda r_{h}^{3}}{6a\pi r_{h}-12\pi r_{h}^{2}}.
\end{equation}
In the extended phase space of AdS BHs, the cosmological constant is related to the BH's pressure
\begin{equation}\label{eq3_1_3}
	P=-\frac{\Lambda}{8\pi}.
\end{equation} 
The BH's entropy satisfies with the Bekenstein-Hawking relation
\begin{equation}\label{eq3_1_4}
	S=\frac{A}{4}=\pi r_{h}^{2},
\end{equation}
where $A=\!\displaystyle\int\!\!\!\!\int\sqrt{ g_{\theta\theta}g_{\phi\phi}}\mathrm{d}\theta\mathrm{d}\phi=4\pi r_{h}^{2}$ is the surface area of the BH's event horizon. The above functions are helpful for deriving the thermodynamic first law. But, when the energy-momentum tensor $T_{0}^{0}$ includes the BH's mass (as seen in Eq.~\ref{eq2_4}), the conventional first law of thermodynamics $\mathrm{d}M=T\mathrm{d}S+V\mathrm{d}P$ will be violated~\cite{thermo_4,thermo_5,rev2}. It's easy to verify
\begin{equation}\label{eq3_1_5}
	T\neq\left(\frac{\partial M}{\partial S}\right)_{P}.
\end{equation}
The modified first law is written as
\begin{equation}\label{eq3_1_6}
	\mathrm{d}\mathcal{M}=W\mathrm{d}M=T\mathrm{d}S+V\mathrm{d}P+\Phi_{a}\mathrm{d}a,
\end{equation}
where $W$ is the correction function,
\begin{equation}\label{eq3_1_7}
	W=1+\int_{r_{h}}^{+\infty}4\pi r^{2}\frac{\partial T_{0}^{0}}{\partial M}\mathrm{d}r=1-\frac{a}{2r_{h}}+\mathcal{O}\left(a^{3}\right).
\end{equation}
Using the modified first law, one get
\begin{equation}\label{eq3_1_8}
	T=W\left(\frac{\partial M}{\partial S}\right)_{P,a},
\end{equation}
and the BH's thermodynamic volume
\begin{equation}\label{eq3_1_9}
	V=W\left(\frac{\partial M}{\partial P}\right)_{S,a}=\frac{4\pi r_{h}^{3}}{3}.
\end{equation}
And
\begin{equation}\label{eq3_1_10}
	\Phi_{a}=W\left(\frac{\partial M}{\partial a}\right)_{S,P}=\frac{-3r_{h}+\Lambda r_{h}^{3}}{6a-12r_{h}}
\end{equation}
could be interpreted as "non-commutative potential". Treating the noncommutative parameter as a thermodynamic variable is not merely a formal requirement but also facilitates the derivation of the BH's Smarr relation.

For BH's mass function $M=M\left(S,P,a\right)$, performing a scale transformation $r_{h}\rightarrow kr_{h}$ on the system, one could get
\begin{equation}\label{eq3_1_11}
	k^{d_{M}}M=M\left(k^{d_{S}}S,k^{d_{P}}P,k^{d_{a}}a\right),
\end{equation}
where $d_{M}$, $d_{S}$, $d_{P}$ and $d_{a}$ are scale dimensions of $M$, $S$, $P$ and $a$. Taking the derivative of both sides of the above equation with respect to $k$, one obtain
\begin{equation}\label{eq3_1_12}
	d_{M}k^{d_{M}-1}M=M_{S}d_{S}k^{d_{S}-1}S+M_{P}d_{P}k^{d_{P}-1}P+M_{a}d_{a}k^{d_{a}-1}a,
\end{equation}
where $M_{S}=\frac{\partial M}{\partial S}$. The same applies to $P$ and $a$. One could set $k=1$ and derive
\begin{equation}\label{eq3_1_13}
	d_{M}M=M_{S}d_{S}S+M_{P}d_{P}P+M_{a}d_{a}a.
\end{equation}
According to Eqs.~\ref{eq2_5}, \ref{eq3_1_3}, \ref{eq3_1_4}, $d_{M}=1,d_{S}=2,d_{P}=-2$, $d_{a}=1$. So one finally obtain
\begin{equation}\label{eq3_1_14}
	M=2\left(M_{S}S-M_{P}V\right)+M_{a}a.
\end{equation}
With the help of Eqs.~\ref{eq3_1_8}, \ref{eq3_1_9}, \ref{eq3_1_10}, one have
\begin{equation}\label{eq3_1_15}
	WM=2\left(TS-PV\right)+\Phi_{a}a.
\end{equation}
It is clear that when $W=1$, the conventional Smarr relation appears. This relation is general. One only needs to fully consider quantities with scale dimensions and derive the Smarr relation using the same method. The general Smarr relation is follows
\begin{equation}\label{eq3_1_16}
	WM=2\left(TS-PV\right)+d_{y^{i}}\Phi_{y^{i}}y^{i},
\end{equation}
Here, $y^{i}$ represents the $i$-th thermodynamic variable other than entropy and pressure, $d_{y^{i}}$ denotes its scale dimension, $\Phi_{y^{i}}=W\frac{\partial M}{\partial y^{i}}$, and repeated indices imply summation.

Strictly speaking, according to Eq.~\ref{eq3_1_7}, it should be considered that the first law of thermodynamics is only violated when the energy-momentum tensor of the spacetime outside the event horizon is affected by the BH's mass. In fact, the energy-momentum tensor of a Schwarzschild BH is also related to the BH's mass, that is
\begin{equation}\label{eq3_1_17}
	T_{0}^{0}=-M\delta^{3}\left(x\right),
\end{equation}
where $\delta^{3}\left(x\right)$ is the Dirac delta function in three-dimensional space.

It should be noted that for the modified first law of thermodynamics, in general, Eq.~\ref{eq3_1_6} is not an exact differential form (see the proof in Appendix.~\ref{App1}). In other words, there does not exist a function $\mathcal{M}$ satisfying Eq.~\ref{eq3_1_6}. This issue will appear again when discussing the Gibbs free energy.

Now, we investigate the Smarr relation using the Komar integral~\cite{komar1,komar2}. A Killing vector in a static spherically symmetric spacetime is given by $\xi=\partial_{t}$, $\xi^{\mu}=\left(1,0,0,0\right)$. To construct the Komar integral relation in the presence of a nonzero cosmological constant, it is necessary to introduce the Killing potential $\omega^{\mu\nu}$~\cite{firstlaw,komar2,komar3,komar4}, which satisfies
\begin{equation}\label{eq3_1_18}
	\nabla_{\mu}\omega^{\mu\nu}=\xi^{\nu}.
\end{equation}
The Killing potential is not unique. In this work, we take the form
\begin{equation}\label{eq3_1_19}
	\omega^{rt}=-\omega^{tr}=\frac{r}{3}.
\end{equation}
One could integrate the Killing equation $\nabla_{\mu}\nabla^{\mu}\xi^{\nu}=-R_{\mu}^{\nu}\xi^{\mu}$ over a three-dimensional hypersurface $\Sigma$
\begin{equation}\label{eq3_1_20}
	\int_{\Sigma}\nabla_{\mu}\nabla^{\mu}\xi^{\nu}\mathrm{d}\Sigma_{\nu}=-\int_{\Sigma}R_{\mu}^{\nu}\xi^{\mu}\mathrm{d}\Sigma_{\nu},
\end{equation}
where $\mathrm{d}\Sigma_{\nu}$ is the surface elements of $\Sigma$. According to the Einstein equation Eq.~\ref{eq2_1}
\begin{equation}\label{eq3_1_21}
	R_{\mu}^{\nu}=\frac{1}{2}\delta_{\mu}^{\nu}R-\delta_{\mu}^{\nu}\Lambda+8\pi T_{\mu}^{\nu}.
\end{equation}
In non-commutative Lorentzian spacetime, one could obtain
\begin{equation}\label{eq3_1_22}
	R_{\mu}^{\nu}\xi^{\mu}=\Lambda\nabla_{\mu}\omega^{\mu\nu}+8\pi T_{\mu}^{\nu}\xi^{\mu}.
\end{equation}
So one have
\begin{equation}\label{eq3_1_23}
	\int_{\Sigma}\nabla_{\mu}\nabla^{\mu}\xi^{\nu}\mathrm{d}\Sigma_{\nu}+\Lambda\nabla_{\mu}\omega^{\mu\nu}\mathrm{d}\Sigma_{\nu}+8\pi T_{\mu}^{\nu}\xi^{\mu}\mathrm{d}\Sigma_{\nu}=0.
\end{equation}
By applying the Gauss theorem, one can derive
\begin{equation}\label{eq3_1_24}
	\int_{\partial\Sigma}\mathrm{d}\Sigma_{\mu\nu}\left(\nabla^{\mu}\xi^{\nu}+\Lambda\omega^{\mu\nu}\right)+\int_{\Sigma}\mathrm{d}\Sigma_{\nu}8\pi T_{\mu}^{\nu}\xi^{\mu}=0,
\end{equation}
where $\partial\Sigma$ is the boundary of $\Sigma$ and $\mathrm{d}\Sigma_{\mu\nu}$ is the surface elements of $\partial\Sigma$. By choosing $\Sigma$ as the entire space outside the horizon, then $\partial\Sigma=S^{2}_{r_{h}}\cup S^{2}_{\infty}$, where $S^{2}_{r_{h}}$ represents the horizon sphere and $S^{2}_{\infty}$ corresponds to the sphere at infinity. Specifically, for a standard Schwarzschild spacetime, no matter exists outside the horizon, which leads to the Komar integral relation in~\cite{firstlaw,komar2}
\begin{equation}\label{eq3_1_25}
	\int_{\partial\Sigma}\mathrm{d}\Sigma_{\mu\nu}\left(\nabla^{\mu}\xi^{\nu}+\Lambda\omega^{\mu\nu}\right)=0.
\end{equation}
For non-commutative Lorentzian BH, due to the influence of the noncommutative parameter $a$, the mass distribution of the BH can be regarded as a fluid distributed throughout the entire spacetime. Now, we proceed with the computation of the Komar integral. The area element of a two-dimensional sphere is given by $\mathrm{d}\Sigma_{tr}=-\mathrm{d}\Sigma_{rt}=-r^{2}\mathrm{d}\Omega/2$, where $\mathrm{d}\Omega$ is the solid angle differential element. For the spherical surface at infinity, one have
\begin{equation}\label{eq3_1_26}
	\int_{S^{2}_{\infty}}\mathrm{d}\Sigma_{\mu\nu}\left(\nabla^{\mu}\xi^{\nu}+\Lambda\omega^{\mu\nu}\right)=4\pi M.
\end{equation}
One could calculate and obtain the nonzero components of $\nabla^{\mu}\xi^{\nu}$
\begin{equation}\label{eq3_1_27}
	\nabla^{t}\xi^{r}=-\nabla^{r}\xi^{t}=-\frac{\partial_{r}f}{2}.
\end{equation}
For the horizon sphere, the first term of Eq.~\ref{eq3_1_25} is
\begin{equation}\label{eq3_1_28}
	\int_{S^{2}_{r_{h}}}\mathrm{d}\Sigma_{\mu\nu}\nabla^{\mu}\xi^{\nu}=\int_{S^{2}_{r_{h}}}r^{2}\mathrm{d}\Omega\nabla^{t}\xi^{r}=2\pi TA=8\pi TS.
\end{equation}
The second term of Eq.~\ref{eq3_1_25} is
\begin{equation}\label{eq3_1_29}
	\int_{S^{2}_{r_{h}}}\mathrm{d}\Sigma_{\mu\nu}\Lambda\omega^{\mu\nu}=-8\pi P\int_{S^{2}_{r_{h}}}\omega^{tr}r^{2}\mathrm{d}\Omega=-8\pi PV.
\end{equation}
The second term of Eq.~\ref{eq3_1_24} is exactly the charge associated with the conserved current $J^{\mu}=T^{\mu\nu}\xi_{\nu}$. One could substitute Eq.~\ref{eq2_4} and derive
\begin{equation}\label{eq3_1_30}
	\int_{\Sigma}\mathrm{d}\Sigma_{\nu}8\pi T_{\mu}^{\nu}\xi^{\mu}=-\frac{4\pi aM}{r_{h}}=-8\pi\Phi_{a}a.
\end{equation}
Combining Eqs.~\ref{eq3_1_24}, \ref{eq3_1_26}, \ref{eq3_1_28}, \ref{eq3_1_29}, \ref{eq3_1_30}, we obtain
\begin{equation}\label{eq3_1_31}
	M=2TS-2PV+2\Phi_{a}a.
\end{equation}
This conclusion is consistent with Eq.~\ref{eq3_1_15}, because it's not hard to verify
\begin{equation}\label{eq3_1_32}
	WM=M-\Phi_{a}a.
\end{equation}
It is important to note that, unlike in the RN spacetime, the non-commutative parameter $a$ is not suggested as an independent conserved charge, as it is coupled to the BH's mass $M$. In this case, the conserved charge of the spacetime is still more appropriately suggested as the energy of the gravitational field. From Eq.~\ref{eq3_1_26}, it follows that the total energy of the gravitational field remains equal to the BH's mass $M$.

\subsection{Critical point}\label{Sect3_2}
\begin{figure}[htbp]
	\centering
	\includegraphics[width=0.46\textwidth]{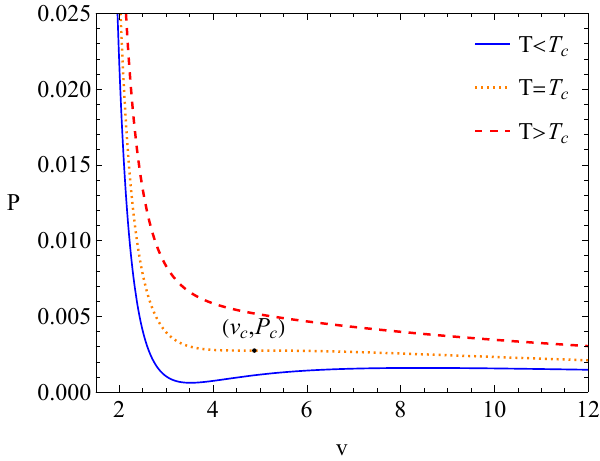}
	\caption{The graph of $P\left(v\right)$ of Schwarzschild-AdS BH in non-commutative geometry. We set $a=1$, and $T=0.8T_{c}$, $T=T_{c}$, $T=1.3T_{c}$ respectively. The black point is the critical point.}\label{Pv}
\end{figure}

To investigate the criticality, the specific volume $v=2r_{h}$ is introduced~\cite{RNcriticality,v}. By using this relation, Eq.~\ref{eq3_1_2} and Eq.~\ref{eq3_1_3} will give the equation of state
\begin{equation}\label{eq3_2_1}
	P=\frac{6a-3v-6a\pi Tv+6\pi T v^{2}}{6\pi v^{3}-8a\pi v^{2}}.
\end{equation}
The critical point meets
\begin{equation}\label{eq3_2_2}
	\frac{\partial P}{\partial v}=\frac{\partial^{2} P}{\partial v^{2}}=0,
\end{equation}
which leads to
\begin{equation}\label{eq3_2_3}
	v_{c}=4.8831a,~~~
	T_{c}=\frac{0.036403}{a},~~~
	P_{c}=\frac{0.0027338}{a^{2}}.
\end{equation}
The critical ratio is
\begin{equation}\label{eq3_2_4}
	\frac{P_{c}v_{c}}{T_{c}}=0.36671.
\end{equation}
This value is smaller than that of the Van der Waals system (0.375)~\cite{RNcriticality}. The graph of $P\left(v\right)$ is plotted in Fig.~\ref{Pv}.
As seen in the figure, there is only one critical point for a fixed parameter $a$. For an isothermal process $T<T_{c}$, the small BH-large BH phase transition occurs. The part where the pressure increases is unstable, and useful solution is to introduce an isobaric curve, which represents the process of the phase transition. The position of this isobaric curve is determined by Maxwell's area law~\cite{RNcriticality,thermo_4,maxareapv1,maxareapv2}
\begin{equation}\label{eq3_2_5}
	\oint V\mathrm{d}P=0.	
\end{equation}

\begin{figure}[htbp]
	\centering
	\includegraphics[width=0.46\textwidth]{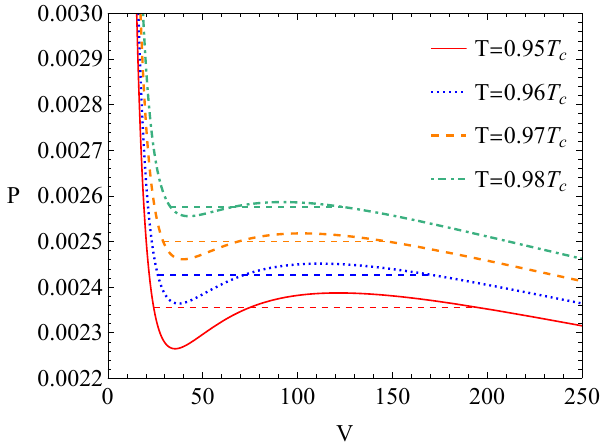}
	\caption{Maxwell's area law in the $\left(V,P\right)$ phase graph. The dashed lines represent the phase transitions. We set $a=1$.}\label{Maxwellarea}
\end{figure}

Fig.~\ref{Maxwellarea} illustrates the result of Maxwell's area law. And it is clear that with the increase of temperature, the phase transition pressure also increases.

By defining
\begin{equation}\label{eq3_5_1}
	p=\frac{P}{P_{c}},~~~
	\tau=\frac{T}{T_{c}},~~~
	\nu=\frac{v}{v_{c}},
\end{equation}
one could get the equation of corresponding state
\begin{equation}\label{eq3_5_2}
	p=\frac{a_{0}+a_{1}\nu+a_{2}\nu\tau+a_{3}\nu^{2}\tau}{a_{4}\nu^{2}+a_{5}\nu^{3}},
\end{equation}
where $a_{0}, a_{1}, a_{2}, a_{3}, a_{4}, a_{5}$ are nonzero constants. To calculate the critical exponents, a transformation is also performed
\begin{equation}\label{eq3_5_3}
	t=\frac{T}{T_{c}}-1=\tau-1,~~~
	w=\frac{V}{V_{c}}-1=\nu^{3}-1.
\end{equation}
After performing this transformation, the equation of state is expanded as a series, yielding
\begin{equation}\label{eq3_5_4}
	p=1+c_{10}t+c_{11}tw+c_{03}w^{3}+\mathcal{O}\left(tw^{2},tw^{3},w^{4}\right),
\end{equation}
where $c_{10}, c_{11}, c_{03}$ are nonzero constants.

There are four critical exponents~\cite{exp1}:
\begin{align}
	C_{V}=T\left(\frac{\partial S}{\partial T}\right)_{V}\propto\left|t\right|^{-\alpha},~\label{eq3_5_5}\\
	\Delta V\propto\left|t\right|^{\beta},~\qquad\quad\label{eq3_5_6}\\
	\kappa_{T}=-\frac{1}{V}\left(\frac{\partial V}{\partial P}\right)_{T}\propto\left|t\right|^{-\gamma},\label{eq3_5_7}\\
	\left|P-P_{c}\right|\propto\left|V-V_{c}\right|^{\delta},~\quad\label{eq3_5_8}
\end{align}
where $C_{V}$ is the isochoric heat capacity. $\Delta V$ is the phase transition volume, the change of the thermodynamic volume during the phase transition. $\kappa_{T}$ is the isothermal compressibility. Eq.~\ref{eq3_5_8} describes the variation of the pressure with the thermodynamic volume at the critical temperature $T=T_{c}$.

Considering $S=\frac{3^{2/3}\pi^{1/3}V^{2/3}}{4^{2/3}}$, the isochoric heat capacity is
\begin{equation}\label{eq3_3_1}
	C_{V}=T\left(\frac{\partial S}{\partial T}\right)_{V}=0,
\end{equation}
which gives $\alpha=0$.

In order to calculate the phase transition volume, Maxwell's area law is necessary. Eq.~\ref{eq3_2_5} yields
\begin{equation}\label{eq3_5_9}
	\int_{w_{1}}^{w_{2}}\left(w+1\right)\frac{\mathrm{d}p}{\mathrm{d}w}\mathrm{d}w=0.
\end{equation}
With the condition $p\left(w_{1}\right)=p\left(w_{2}\right)$, the solution could be obtained
\begin{equation}\label{eq3_5_10}
	w_{2}=-w_{1}=\sqrt{\frac{c_{11}t}{c_{03}}}.
\end{equation}
This indicates $\beta=\frac{1}{2}$.

Eq.~\ref{eq3_5_7} could be calculated as
\begin{equation}\label{eq3_5_11}
	\kappa_{T}=-\frac{1}{\left(w+1\right)P_{c}}\left(\frac{\partial w}{\partial p}\right)_{T}=-\frac{1}{c_{11}P_{c}t}+\mathcal{O}\left(w\right),
\end{equation}
which yields $\gamma=1$.

When $T=T_{c}$ ($t=0$),
\begin{equation}\label{eq3_5_12}
	p=1+c_{03}w^{3},
\end{equation}
which shows $\delta=3$.

It is concluded that four critical exponents of the Schwarzschild-AdS BH in non-commutative geometry are the same as those of the Van der Waals system~\cite{RNcriticality}. These four exponents conform to the Griffiths, Rushbrooke and Widom functions~\cite{maxareapv1,exp1,exp2}
\begin{align}
	&\text{Griffiths}\!:~\alpha+\beta\left(\delta+1\right)-2=0,\label{eq3_5_13}\\
	&\text{Griffiths}\!:~\gamma\left(\delta+1\right)+\left(\alpha-2\right)\left(\delta-1\right)=0,\label{eq3_5_14}\\
	&\text{Rushbrooke}\!:~\alpha+2\beta+\gamma-2=0,\label{eq3_5_15}\\
	&\text{Widom}\!:~\gamma-\beta\left(\delta-1\right)=0.\label{eq3_5_16}
\end{align}
\subsection{Isobaric heat capacity}\label{Sect3_3}
\begin{figure}[htbp]
	\centering
	\includegraphics[width=0.46\textwidth]{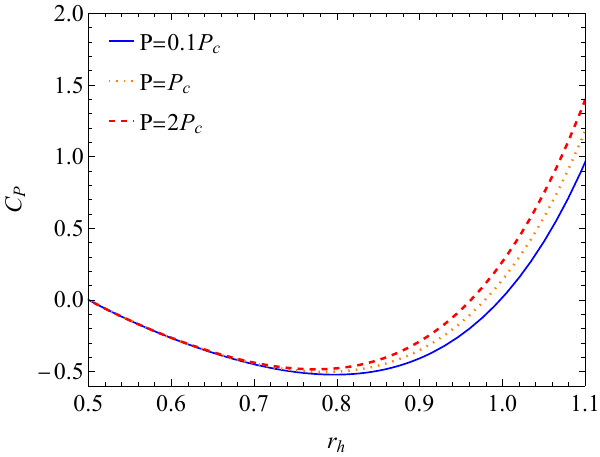}
	\caption{The region where the heat capacity $C_{P}$ is negative when $r_{h}$ is too small. We set $a=1$.}\label{cpn}
\end{figure}

\begin{figure}[htbp]
	\centering
	\includegraphics[width=0.46\textwidth]{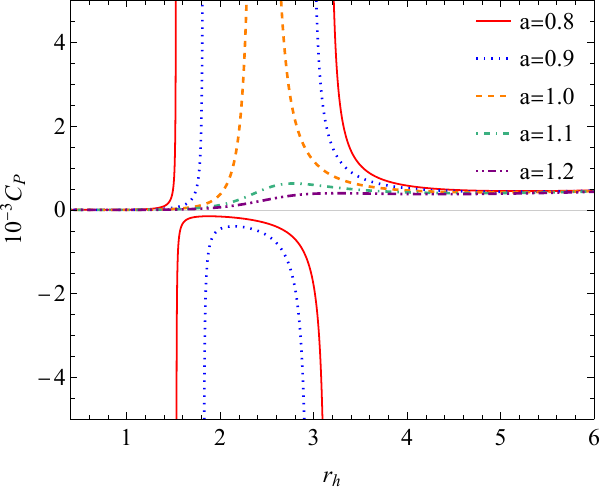}
	\caption{The heat capacity for different non-commutative parameters $a$. We set pressure $P=P_{c}\left(a=1\right)$.}\label{cpa}
\end{figure}

The isobaric heat capacity is
\begin{widetext}
	\begin{equation}\label{eq3_3_2}
		C_{P}=T\left(\frac{\partial S}{\partial T}\right)_{P}=
		\frac{2\pi r_{h}^2\left(2r_{h}-a\right)\left(3r_{h}+24P\pi r_{h}^3-16aP\pi r_{h}^2-3a\right)}{6 r_{h}^2 \left(8 P\pi r_{h}^2-1\right)+a^2\left(16 P\pi r_{h}^2-3\right)+a\left(12 r_{h}-48P\pi r_{h}^3\right)}
	\end{equation}
	\begin{center}
		\begin{figure}[htbp]
			\centering
			\includegraphics[width=1\textwidth]{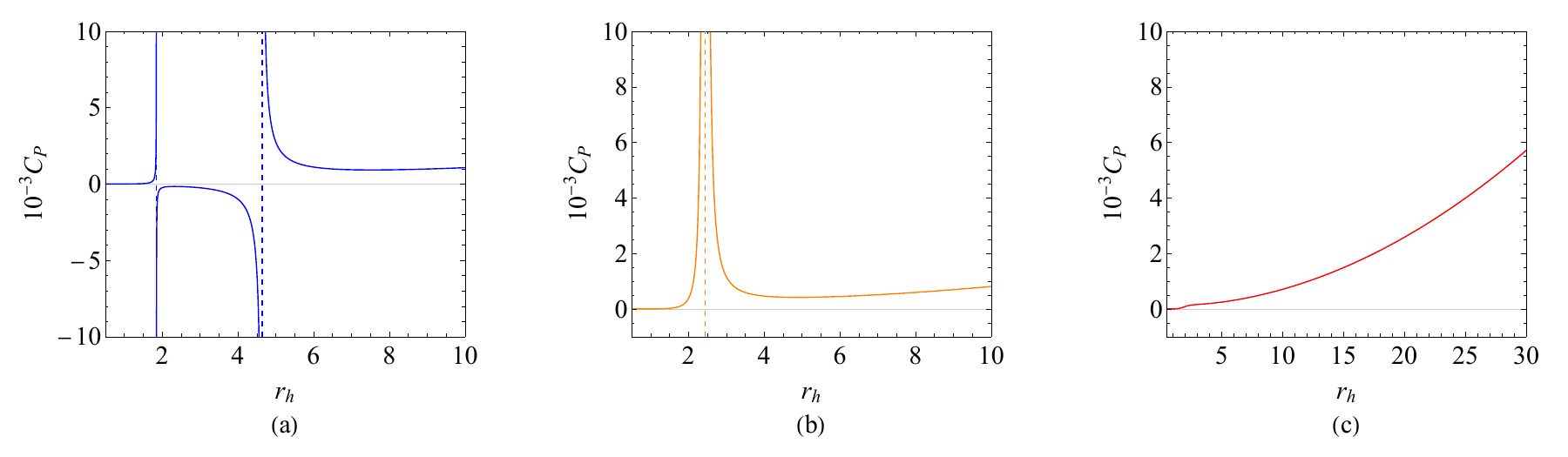}
			\caption{The heat capacity $C_{P}$ as a function of $r_{h}$. We set (a) $a=1, P=0.5P_{c}$, (b) $a=1, P=P_{c}$ and (c) $a=1, P=2P_{c}$.}\label{cp1}
		\end{figure}
	\end{center}
\end{widetext}
Fig.~\ref{cp1} shows the heat capacity as a function of $r_{h}$. When $P<P_{c}$, there exist two different points satisfying $\frac{\partial T}{\partial S}=0$ (marked by the dashed lines), resulting in the divergence of $C_{P}$. Meanwhile, there is an unstable region with negative heat capacity (represented by the segment in Fig.~\ref{Pv} and Fig.~\ref{Maxwellarea} where $P$ increases monotonically with $V$). As mentioned earlier, this unstable region can be eliminated using Maxwell's area law. As the pressure increases, phase transition points gradually combine into one at the critical point. At the critical point, the heat capacity still diverges to infinity, but the sign of the heat capacity no longer changes. Study also showed that when the radius of horizon is small, the heat capacity also could be negative. This interesting phenomenon is demonstrated in Fig.~\ref{cpn}. This negative region is caused by the BH's temperature $T<0$. From a thermodynamic perspective, it indicates that AdS-BH with a extremely small event horizon radius is unstable.
This negative region is
\begin{equation}\label{eq3_3_3}
	\frac{a}{2}<r_{h}<r_{min},
\end{equation}
where $r_{min}$ is the positive root of the equation
\begin{equation}\label{eq3_3_4}
	-3a+3r_{h}-16aP\pi r_{h}^{2}+24P\pi r_{h}^{3}=0.    
\end{equation}
We also investigated the influence of the non-commutative parameter $a$. As seen in Fig.~\ref{cpa}, with the increase of $a$, the distance between two phase transition points gradually shortens and eventually turns into one at the critical point. As $a$ continues to increase, the phase transition disappears.

\subsection{Gibbs free energy and zeroth-order phase transition}\label{Sect3_4}
\begin{figure}[htbp]
	\centering
	\includegraphics[width=0.46\textwidth]{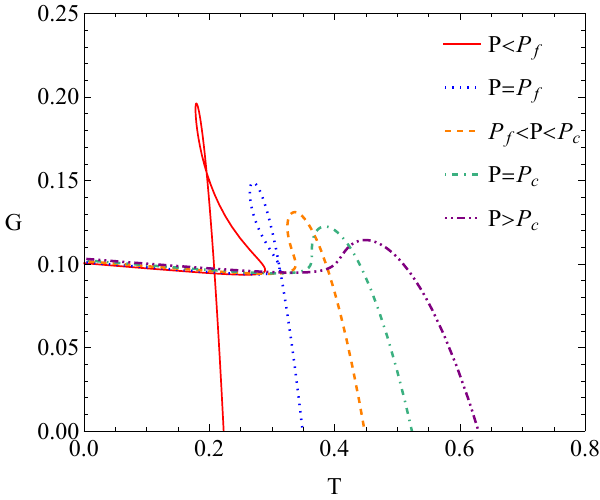}
	\caption{The Gibbs for different pressure $P$. We set $a=0.1$ and $P=0.2P_{c},P_{f},0.75P_{c},P_{c},1.4P_{c}$.}\label{Gibbs}
\end{figure}

\begin{figure}[htbp]
	\centering
	\includegraphics[width=0.46\textwidth]{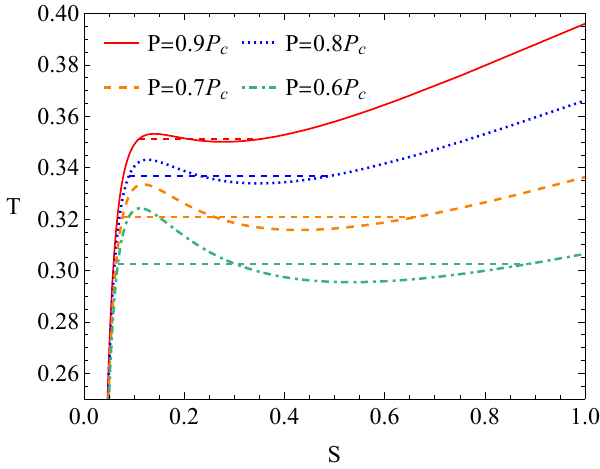}
	\caption{Maxwell's area law in $\left(S,T\right)$ coordinate system. The dashed lines represent the phase transitions. We set $a=0.1$.}\label{MaxwellareaTS}
\end{figure}

\begin{figure}[htbp]
	\centering
	\includegraphics[width=0.46\textwidth]{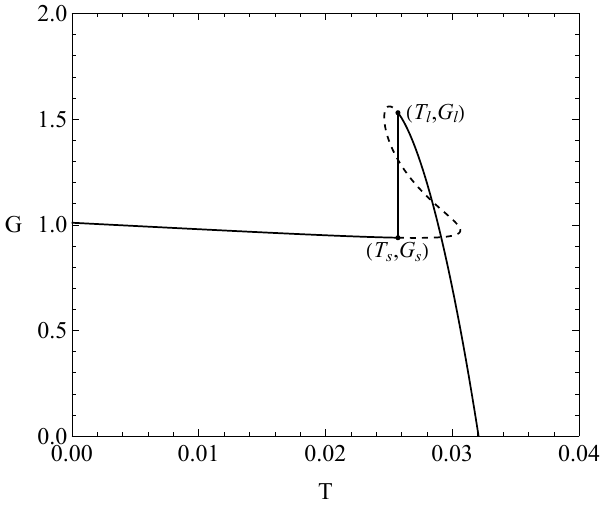}
	\caption{The graph of the Gibbs free energy $G\left(T\right)$ when considering Maxwell's area law. The points $\left(T_{s},G_{s}\right)$ and $\left(T_{l},G_{l}\right)$ represent the small BH and the large BH. We set $a=1$ and $P=0.4P_{c}$.}\label{gzx}
\end{figure}

\begin{figure}[htbp]
	\centering
	\includegraphics[width=0.46\textwidth]{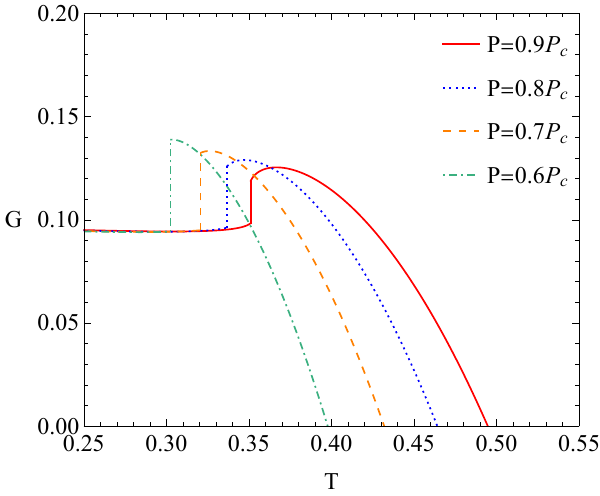}
	\caption{The Gibbs free energy corrected by Maxwell's area law. We set $a=0.1$.}\label{Gibbscorrected}
\end{figure}

In this section, we study the Gibbs free energy. The definition is
\begin{equation}\label{eq3_4_1}
	G=M-TS=\frac{r_{h}\left(3r_{h}-8\pi Pr_{h}^{3}+a\left(3+16\pi Pr_{h}^{2}\right)\right)}{6\left(2r_{h}-a\right)}.
\end{equation}
It is well known in conventional thermodynamics, Gibbs free energy satisfies
\begin{equation}\label{eq3_4_2}
	\mathrm{d}G=-S\mathrm{d}T+V\mathrm{d}P
\end{equation}
For a constant pressure system, the change of the Gibbs free energy during a thermodynamic process is
\begin{equation}\label{eq3_4_3}
	\Delta G=-\int S\mathrm{d}T.
\end{equation}
Considering Maxwell's area law in $\left(S,T\right)$ coordinate system~\cite{maxareats}
\begin{equation}\label{eq3_4_4}
	\oint S\mathrm{d}T=0,
\end{equation}
and the first-order phase transition is an isothermal process, there is
\begin{equation}\label{eq3_4_5}
	\Delta G=-\int S\mathrm{d}T=0.
\end{equation}
The analysis of phase transitions for an isothermal system is similar. This implies that the Gibbs free energy of a thermodynamic system does not change before and after a first-order phase transition. More precisely, there is a conclusion that any isothermal and isobaric process that a system undergoes (which could be an irreversible process) must proceed in the direction of non-increasing Gibbs free energy. Specifically, for a reversible process occurring in an isothermal and isobaric system, the Gibbs free energy remains unchanged. But for the Schwarzschild-AdS BH in non-commutative geometry, there is a issue, since $T\mathrm{d}S+V\mathrm{d}P$ is not an exact form, it follows that $-S\mathrm{d}T+V\mathrm{d}P$ is also not an exact form. In this case, the differential of $G$ is
\begin{equation}\label{eq3_4_6}
	\mathrm{d}G=\left(W^{-1}-1\right)T\mathrm{d}S+W^{-1}V\mathrm{d}P-S\mathrm{d}T+W^{-1}\Phi_{a}\mathrm{d}a.
\end{equation}
The Gibbs free energy is plotted in Fig.~\ref{Gibbs}.
In the locally monotonic region of the $G\left(T\right)$, the differential form of $G$ can be rewritten as follows
\begin{widetext}
	\begin{equation}\label{eq3_4_7}
		\mathrm{d}G=\left(-S+\left(W^{-1}-1\right)C_{P}\right)\mathrm{d}T+\left(W^{-1}V-\left(W^{-1}-1\right)C_{P}\left(\frac{\partial T}{\partial P}\right)_{S}\right)\mathrm{d}P,
	\end{equation}
\end{widetext}
which leads to
\begin{equation}\label{eq3_4_8}
	\frac{\partial G}{\partial T}=-S+\left(W^{-1}-1\right)C_{P},
\end{equation}
which is different from $\frac{\partial G}{\partial T}=-S<0$ in conventional thermodynamics, causing that $G\left(T\right)$ may exhibit a monotonic increasing region. When $P<P_{c}$, since the function $T\left(S\right)$ is not monotonic, the Gibbs free energy becomes locally multivalued. When $P>P_{c}$, the Gibbs free energy becomes a single-valued function of $T$ and $P$. Specifically, there exists a pressure $P_{f}$ such that when $P<P_{f}$, the curve $G\left(T\right)$ exhibits self-intersection. It can be obtained that
\begin{equation}\label{eq3_4_9}
	P_{f}=\frac{0.0012845}{a^{2}},
\end{equation}
which gives a dimensionless constant
\begin{equation}\label{eq3_4_10}
	\frac{P_{f}}{P_{c}}=0.46987.
\end{equation}
However, unlike the thermodynamics of Van der Waals system, the intersection points of the curve $G\left(T\right)$ can not be regarded as the phase transition points. This is because there is a change in Gibbs free energy before and after the transition. The location of the phase transition point should be determined by Maxwell's area law for the condition $P<P_{c}$
\begin{equation}\label{eq3_4_11}
	\begin{cases}
		\raisebox{8pt}{$T\left(S_{1}\right)=T\left(S_{2}\right)$,}\\
		\displaystyle \int_{S_{1}}^{S_{2}}T\mathrm{d}S=T\left(S_{1}\right)\left(S_{2}-S_{1}\right),
	\end{cases}
\end{equation}
where $S_{1}$ and $S_{2}$ are the entropies of the initial and final states during the phase transition. The results are shown in Fig.~\ref{MaxwellareaTS}.
When considering Maxwell's area law, the Gibbs free energy is shown by Fig.~\ref{gzx}.
In this case, the Gibbs free energy for different pressures is plotted in Fig.~\ref{Gibbscorrected}.
As shown in the figure, the vertical section of $G\left(T\right)$ represents the phase transition. The change of the Gibbs free energy during a phase transition is given by Eq.~\ref{eq3_4_6}
\begin{equation}\label{eq3_4_12}
	\Delta G=\int_{S_{1}}^{S_{2}}\left(W^{-1}-1\right)T\mathrm{d}S.
\end{equation}
Here, a peculiar phenomenon occurs: according to Ehrenfest's classification of phase transitions, a discontinuous Gibbs free energy indicates the occurrence of a zeroth-order phase transition. The zeroth-order phase transition of BHs is a very intriguing phenomenon. Studies have shown that the dilaton-electromagnetic coupling interaction could lead to a zeroth-order phase transition~\cite{maxareapv2,zero1,zero2}. However, unlike the physical mechanism that induces zeroth-order phase transitions as described in~\cite{maxareapv2,zero1,zero2}, our research indicates that for a static, spherically symmetric AdS BH, the violation of the conventional first law of thermodynamics would lead to a change in the Gibbs free energy of a BH after undergoing a small BH-large BH phase transition.

\begin{figure}[htbp]
	\centering
	\includegraphics[width=0.46\textwidth]{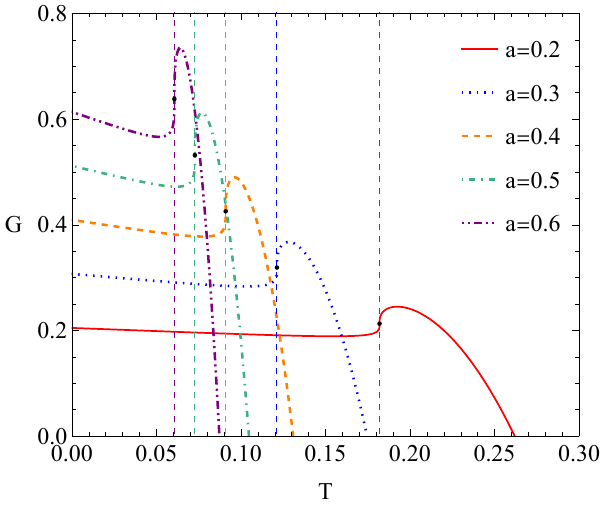}
	\caption{At the critical points, the first-order phase transitions occur. The slope of the Gibbs free energy with respect to temperature diverges to infinity. The black points are the critical points.}\label{firstorder}
\end{figure}

At the critical point, $G\left(T\right)$ is continuous, but the heat capacity $C_{p}$ diverges to infinity, which results in
\begin{equation}
	\frac{\partial G}{\partial T}=-S+\left(W^{-1}-1\right)C_{P}\rightarrow\infty,
\end{equation}
which indicates that the phase transition occurring at the critical point is a first-order phase transition. This phenomenon is shown in Fig.~\ref{firstorder}.

\section{Joule-Thomson process}\label{Sect4}
The Joule-Thomson expansion in BH thermodynamics has been widely studied~\cite{JT1,JT2,JT3,JT4,JT5,JT6,JT7,JT8,JT9,JT10,JT11,JT12,JT13,JT14,JT15,JT16,JT17,JT18}. The Joule-Thomson process is a thermodynamic system's throttling (constant enthalpy) expansion. As previously mentioned, the mass $M$ of an AdS-BH is considered as its enthalpy. Even though the first law has been modified, studying the constant mass process of BHs remains highly significant.

To study the constant mass process, we take $M$ and $r_{h}$ as variables and rewrite the BH's pressure and temperature 
\begin{equation}\label{eq4_1}
	P=\frac{-3aM+6Mr_{h}-3r_{h}^{2}}{8\pi r_{h}^{4}},
\end{equation}
\begin{equation}\label{eq4_2}
	T=\frac{-2aM+3Mr_{h}-r_{h}^{2}}{2\pi r_{h}^{3}}.
\end{equation}
The Joule-Thomson coefficient is
\begin{equation}\label{eq4_3}
	\mu=\left(\frac{\partial T}{\partial P}\right)_{M}=\frac{12aMr_{h}-12Mr_{h}^{2}+2r_{h}^{3}}{6aM-9Mr_{h}+3r_{h}^{2}}.
\end{equation}
A process with $\mu>0$ is referred to as a cooling process, whereas a process with $\mu<0$ is called a heating process. The point where $\mu=0$ is known as the inversion point.

\begin{figure}[htbp]
	\centering
	\includegraphics[width=0.45\textwidth]{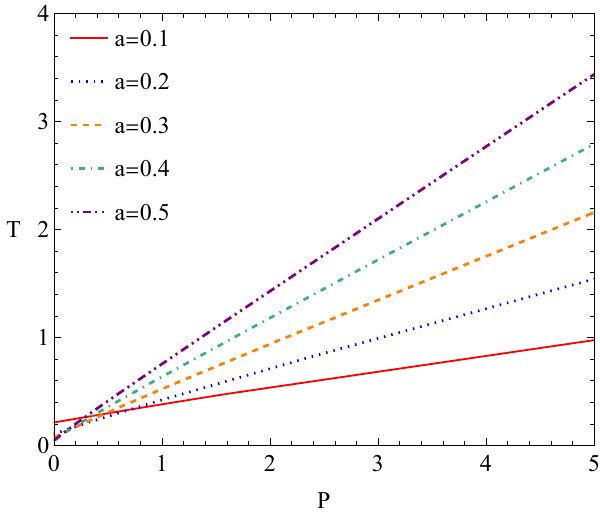}
	\caption{The inversion curves for different non-commutative parameters.}\label{mu0}
\end{figure}

The curve formed by all the inversion points in the $\left(P,T\right)$ coordinates is called the inversion curve. As shown in Fig.~\ref{mu0}, we plotted the inversion curves for different values of the parameter $a$. The figure clearly showed that as the inversion pressure $P_{i}$ increases, the inversion temperature $T_{i}$ also increases. Moreover, when $P_{i}$ is not too small, the $T_{i}$ increases with the increase of non-commutative parameter $a$. In the constant mass curves we will draw next, the inversion curve will divide the $\left(P,T\right)$ phase space into two disconnected regions. The BH's cooling process is located in the upper left of the inversion curve, while the heating process is in the lower right. 

Specifically, when $P_{i}=0$, one have the minimum inversion temperature
\begin{equation}\label{eq4_5}
	T_{i}^{min}=\frac{1}{15a\pi},
\end{equation}
yielding a ratio that is independent of $a$
\begin{equation}\label{eq4_6}
	\frac{T_{i}^{min}}{T_{c}}=0.58294.
\end{equation}
This value lies between the results of the RN-AdS BH (0.5) and the Van der Waals system (0.75). And, interestingly, study had shown that the Joule-Thomson expansion of the Van der Waals systems reveals the presence of a maximum inversion temperature $T_{i}^{max}$~\cite{JT1}, whereas the BH we are studying does not exhibit this phenomenon.

Whether a BH undergoes an inversion point during its constant mass expansion depends on its mass $M$. It can be proven that under the conditions $P>0$, $r_{h}>0$ and $M>0$, the equation $\mu=0$ has a solution if and only if
\begin{equation}\label{eq4_7}
	M>M_{i}^{min}=\frac{25a}{24},
\end{equation}
where $M_{i}^{min}$ is the minimum inversion mass.
\begin{figure}[htbp]
	\centering
	\includegraphics[width=0.49\textwidth]{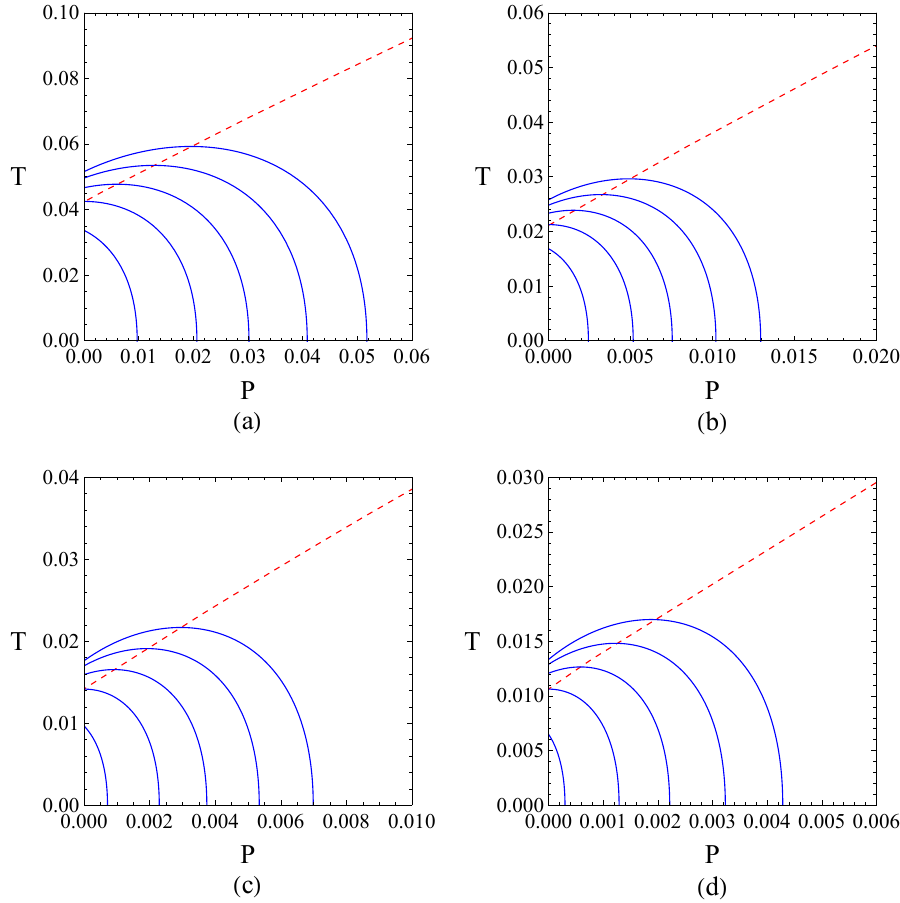}
	\caption{The constant mass curves (blue solid lines) and the inversion curves (red dashed lines) for different parameters $a$. The constant mass curves expand outward as the mass increases. We set (a) $a=0.5,M=0.51,M_{i}^{min}\left(\simeq0.52\right),0.53,0.54,0.55$; (b) $a=1,M=1.02,M_{i}^{min}\left(\simeq1.04\right),1.06,1.08,1.1$; (c) $a=1.5,M=1.52,M_{i}^{min}\left(\simeq1.56\right),1.6,1.64,1.68$; (d) $a=2,M=2.02,M_{i}^{min}\left(\simeq2.08\right),2.14,2.2,2.26$.}\label{JT}
\end{figure}
Fig.~\ref{JT} shows the inversion curves and the families of Joule-Thomson expansion curves for different parameters $a$. It is clearly observed from the figure that BHs with mass less than $M_{i}^{min}$ can not experience an inversion point and will remain in the heating process throughout.

\section{Conclusion and outlook}\label{Sect5}
In this paper, we investigated the thermodynamics of the Schwarzschild-AdS BH within the framework of non-commutative geometry. By solving the Einstein's equations with a Lorentzian distribution and the cosmological constant, we derived a corrected Schwarzschild solution and studied its thermodynamics. Our analysis confirmed that the explicit inclusion of the BH's mass in the energy-momentum tensor disrupts the conventional first law of BH thermodynamics.

The study showed that the BH has a critical point, with a critical ratio of 0.36671. As the non-commutative parameter increases, the phase transition process shortens and eventually disappears after the critical point. Additionally, our calculation revealed that the BH exhibits the same critical exponents as those of the Van der Waals system, suggesting a deeper connection between BH thermodynamics and classical thermodynamic systems.

Furthermore, the violation of the conventional first law caused a sudden change of Gibbs free energy with respect to temperature during a small BH-large BH phase transition. This novel phenomenon indicated a occurrence of the zeroth-order phase transition.

Finally, we analyzed the Joule-Thomson expansion and identified the existence of inversion points, which depend on whether the BH's mass exceeds the critical minimum inversion mass. We also derived the minimum inversion temperature, adding to the understanding of BH thermodynamic behavior under non-commutative geometry. So far, there are no existing works systematically discussing the thermodynamic properties of Schwarzschild-AdS BHs with Lorentzian or Gaussian distributions, such as their first law, phase transitions and criticality, Joule-Thomson effect, etc. We hope that our work could provide more valuable insights for the theoretical study of non-commutative geometry.

In future work, we aim to further investigate the higher-order effects of non-commutative geometry on the Schwarzschild spacetime, as Eq.~\ref{eq2_5} is an approximation retained up to the order of $\mathcal{O}\left(a\right)$. On the other hand, since the BH's mass explicitly appears in the energy-momentum tensor, the conventional first law of thermodynamics is violated, resulting in that differential forms $T\mathrm{d}S+V\mathrm{d}P$ and $-S\mathrm{d}T+V\mathrm{d}P$ are not exact forms. For thermodynamic quantities such as enthalpy and Gibbs free energy, there may be better ways to define these quantities, which would be a valuable topic for future research.

\section*{Conflicts of interest}
The authors declare that there are no conflicts of interest regarding the publication of this paper.

\section*{Acknowledgments}
We want to thank Yu-Cheng Tang and Yu-Hang Feng for their useful suggestions.

\section*{Data availability}
This paper is a theoretical study. No data was used or generated during the research.

\appendix
\section{Differential form $\mathrm{d}\mathcal{M}$ is not an exact form}\label{App1}
\setcounter{equation}{0}
\renewcommand{\theequation}{A.\arabic{equation}}
The modified first law reads
\begin{equation}\label{eqapp1}
	\mathrm{d}\mathcal{M}=T\mathrm{d}S+V\mathrm{d}P.
\end{equation}
Here, we have not explicitly written out the differential terms of the non-commutative parameter $\mathrm{d}a$, as this does not affect the proof. This form is exact if and only if
\begin{equation}\label{eqapp2}
	\left(\frac{\partial T}{\partial P}\right)_{S}=\left(\frac{\partial V}{\partial S}\right)_{P}.
\end{equation}
By substituting
\begin{equation}\label{eqapp3}
	T=W\left(\frac{\partial M}{\partial S}\right)_{P},~~~
	V=W\left(\frac{\partial M}{\partial P}\right)_{S}.
\end{equation}
into Eq.~\ref{eqapp2}, one get
\begin{equation}\label{eqapp4}
	\left(\frac{\partial W}{\partial P}\right)_{S}\left(\frac{\partial M}{\partial S}\right)_{P}=\left(\frac{\partial W}{\partial S}\right)_{P}\left(\frac{\partial M}{\partial P}\right)_{S}.
\end{equation}
By using the chain rule for composite functions $W\left(M,S\right)=W\left(M\left(S,P\right),S\right)$, the above formula gives
\begin{equation}\label{eqapp5}
	\left(\frac{\partial W}{\partial S}\right)_{M}\left(\frac{\partial M}{\partial P}\right)_{S}=0.
\end{equation}
With the relation $S=\pi r_{h}^{2}$, it is derived that
\begin{equation}\label{eqapp6}
	\left(\frac{\partial W}{\partial r_{h}}\right)_{M}=0.
\end{equation}
On the other hand, $W$ is
\begin{equation}\label{eqapp7}
	W\left(M,r_{h}\right)=1+\int_{r_{h}}^{+\infty}4\pi r^{2}\frac{\partial T_{0}^{0}}{\partial M}\mathrm{d}r.
\end{equation}
So Eq.~\ref{eqapp6} can be simplified as
\begin{equation}\label{eqapp8}
	\frac{\partial T_{0}^{0}}{\partial M}\left(M,r_{h}\right)=0.
\end{equation}
However, this condition generally does not hold unless $\frac{\partial T_{0}^{0}}{\partial M}=0$, that is $W=1$.

So $\mathrm{d}\mathcal{M}$ has been proven to not be an exact form. There is another method to prove this fact. If there is a scalar function $\mathcal{M}$ satisfying Eq.~\ref{eq3_1_6}, one could obtain its Smarr relation
\begin{equation}\label{eqapp9}
	\mathcal{M}=2\left(TS-PV\right)+\Phi_{a}a.
\end{equation}
Then one could get
\begin{equation}\label{eqapp10}
	\mathcal{M}=WM,
\end{equation}
which leads to
\begin{equation}\label{eqapp11}
	\mathrm{d}\mathcal{M}=M\mathrm{d}W+W\mathrm{d}M.
\end{equation}
On the other hand, we require that $\mathrm{d}\mathcal{M}=W\mathrm{d}M$. It gives
\begin{equation}\label{eqapp12}
	\mathrm{d}W=0.
\end{equation}
This doesn't hold unless $W$ is a constant.

Form $-S\mathrm{d}T+V\mathrm{d}P$ is not exact either, because if there exists a function $\mathcal{G}$ satisfying $\mathrm{d}\mathcal{G}=-S\mathrm{d}T+V\mathrm{d}P$, then one can construct $\mathcal{M}=\mathcal{G}+TS$ to meet $\mathrm{d}\mathcal{M}=T\mathrm{d}S+V\mathrm{d}P$, which causes a contradiction.

\bibliography{paper}

\end{document}